# Stability analysis of solutions in the helicoidal Peyrard - Bishop model of DNA molecule[*]

**Anna Batova[†]**

Laboratory of radiation biology, Joint Institute for Nuclear Research, 141980 Dubna, Russia

**Dragana Ranković[‡]**

Farmaceutski fakultet, Univerzitet u Beogradu, 11221 Beograd, Serbia

**Slobodan Zdravković[§]**

Institut za Nuklearne Nauke Vinča, Laboratorija za Atomsku Fiziku (040), 11001 Beograd, Serbia

ABSTRACT

We use the helicoidal Peyrard-Bishop model of DNA in the current work. We solve a dynamical equation of motion using a continuum approximation, resulting in kink-solitary waves that travel along the chain. We demonstrate that, whereas supersonic kink solitons are not stable, subsonic ones are. Moreover, we demonstrate the importance of viscosity by showing that no wave is stable in the absence of viscosity.

[*] This work has been supported by the Project within the Cooperation Agreement between the JINR, Dubna, Russian Federation, and the Ministry of Science, Technological Development, and Innovation of the Republic of Serbia: Solitons and chaos in the nonlinear dynamics of biomolecules. This research was also funded by the Ministry of Science, Technological Development, and Innovation, Republic of Serbia, through a grant agreement with the University of Belgrade – Faculty of Pharmacy No. 451-03-47/2023-01/200161.

[†] e-mail address: batova@jinr.ru

[‡] e-mail address: dragana.rankovic@pharmacy.bg.ac.rs

[§] e-mail address: szdjidji@vin.bg.ac.rs

## 1. Introduction

Numerous attempts have been made to establish suitable models to explain the intricate dynamics of DNA molecule [1-3]. The first nonlinear model was presented in 1980 [4], suggesting that nonlinearity focuses the vibration energy of DNA into localised soliton-like excitations. DNA is a double helix composed of nucleotide pairs. Harmonic potential energies can be used to mimic the strong covalent chemical bonds that bind nucleotides belonging to the same strand. On the other hand, the nucleotides belonging to the opposite strands interact via weak hydrogen bonds. Such interactions require anharmonic potentials to be modelled, and the most common example is the Morse potential, used here as well as in many papers dealing with DNA [5-9]. Thus, nonlinear effects in DNA, like in all molecular biological systems, originate from weak interactions.

The structure of the paper is as follows: We provide a brief explanation of the helicoidal Peyrard-Bishop (HPB) model of DNA in Section 2. Following a continuum approximation and Ref. [10], we show that a crucial dynamical equation of motion has two solutions if viscosity is taken into consideration and two if it is neglected. We looked at these solutions' stability in Section 3, and Section 4 is devoted to some concluding remarks.

## 2. Helicoidal Peyrard-Bishop model

In this paper, we use the widely recognised helicoidal Peyrard-Bishop (HPB) model for DNA dynamics [5,6]. This is an improved version of the PB model [11], which ignores DNA's helicoidal structure. We here assume that the reader is familiar with the HPB model [5,6,10]. Therefore, we simply provide a very brief summary of this model in the text that follows. Notice that, in some papers, the HPB model is called the Peyrard-Bishop-Dauxois (PBD) model. However, there is a similar model [12-14] known as the PBD one, and consequently, the HPB is the better option.

We start with the following Hamiltonian:

$$H = \sum \left\{ \frac{m}{2}(\dot{u}_n{}^2 + \dot{v}_n{}^2) + \frac{k}{2}[(u_n - u_{n-1})^2 + (v_n - v_{n-1})^2] \right.$$

$$\left. + \frac{K}{2}[(u_n - v_{n+h})^2 + (u_n - v_{n-h})^2] + D[e^{-a(u_n - v_n)} - 1]^2 \right\}, \qquad (1)$$

which describes the HPB model [5,6,10]. Here, $m = 5.4 \cdot 10^{-25}\,\text{kg}$ is the average nucleotide mass, $u_n$ and $v_n$ are the displacements of the nucleotides at position $n$ from their equilibrium positions along the direction of the hydrogen bond, and a dot denotes the first derivative with respect to time. The parameters $k$, $K$, $D$, and $a$ are explained in Refs. [5,6,10]. The first term in Eq. (1) represents kinetic energy, the second one is the harmonic potential mentioned above, the third term describes the helicoidal structure, and the last one is the Morse potential. We assume $h = 5$ because the helix has a helical pitch of about 10 base pairs per turn [15].

It is convenient to introduce new coordinates

$$x_n = (u_n + v_n)/\sqrt{2}, \qquad y_n = (u_n - v_n)/\sqrt{2}, \qquad (2)$$

representing the in-phase and out-of-phase transversal displacements, respectively. This allows us to obtain two decoupled dynamical equations of motion [5,6,10,16]. We derive these equations directly from Eq. (1) and Hamilton's equations. The nonlinear one, relevant for this paper, is

$$m\ddot{y}_n = k(y_{n+1} + y_{n-1} - 2y_n) - K(y_{n+h} + y_{n-h} + 2y_n)$$
$$+ 2\sqrt{2}aD(e^{-a\sqrt{2}y_n} - 1)\, e^{-a\sqrt{2}y_n} - \gamma\dot{y}_n. \qquad (3)$$

The last term in Eq. (3) is a viscosity force, where $\gamma$ is a viscosity coefficient [17–19].

In essence, we are able to employ both semi-discrete and continuum approximations. The first one yields breathers moving along the chain [3,5,6,20,21]. This approach has been used to study DNA-RNA transcription [16,22] and to characterise a local opening as a resonance mode [23,24]. Here, we are using the continuum approximation $y_n(t) \to y(x,t)$, which, together with the series expansions

$$y_{n+1} + y_{n-1} - 2y_n \to \frac{\partial^2 y}{\partial x^2} l^2, \quad (e^{-a\sqrt{2}y_n} - 1)\, e^{-a\sqrt{2}y_n} \to -a\sqrt{2}y + 3a^2 y^2 \quad (4)$$

results in

$$m\frac{\partial^2 y}{\partial t^2} - l^2\left(k - Kh^2\right)\frac{\partial^2 y}{\partial x^2} + Ay - By^2 + \gamma\frac{\partial y}{\partial t} = 0, \qquad (5)$$

where $l = 3.4\,\overset{\circ}{\text{A}}$ is the distance between the two neighbouring nucleotides in the same strand, and $A = 4\left(K + a^2 D\right)$ and $B = 6\sqrt{2}a^3 D$ [10].

It is well known that, for a given wave equation, a travelling wave $y(\xi)$ is a solution that depends on a unified variable $\xi \equiv \kappa x - \omega t$, where $\kappa$ and $\omega$ are constants. The following ordinary differential equation can be simply obtained by introducing a dimensionless function $\psi$ through $y = (A/B)\psi$ [10]

$$\alpha \frac{d^2\psi}{d\xi^2} - \rho \frac{d\psi}{d\xi} + \psi - \psi^2 = 0,\qquad(6)$$

where

$$\alpha = \frac{m\omega^2 - l^2\kappa^2\left(k - Kh^2\right)}{A},\qquad \rho = \frac{\gamma\omega}{A}.\qquad(7)$$

There are numerous methods for resolving Eq. (9). The simplest equation method (SEM) [25–27], modified SEM [28], procedure based on Jacobian elliptic functions [29,30], method of factorization [31,32], exponential function procedure [33,34], standard procedure [17,35], modified extended tanh-function (METHF) method [36–38], etc. are a few of them. Except for the standard procedure and method of factorization, in all other methods the function $\psi$ is expected to be a series of known functions. However, it is also feasible to expand the series in terms of unknown functions [39,40]. Here, we use the METHF method, probably the simplest procedure for representing the series expansion in terms of known functions, and look only for solutions that have physical sense. According to this procedure, we look for the possible solutions to Eq. (6) in the form

$$\psi = a_0 + \sum_{i=1}^{M}\left(a_i\Phi^i + b_i\Phi^{-i}\right),\qquad(8)$$

where the function $\Phi = \Phi(\xi)$ is a solution of the well-known Riccati equation [36-38]

$$\frac{d\Phi}{d\xi} = b + \Phi^2.\qquad(9)$$

A solution to Eq. (9) can be expressed through tangent and tangent hyperbolic. We choose $b_i = 0$ in Eq. (8) to eliminate diverging solutions and

$$\Phi = -\sqrt{-b}\,\tanh\!\left(\sqrt{-b}\,\xi\right),\qquad(10)$$

which holds for $b < 0$ [10]. As $M = 2$ in this case [28], we obtain, using Mathematica, the following two solutions, i.e., the values of the parameters $b$, $a_0$, $a_1$, $a_2$ and $\alpha$:

$$\left.\begin{array}{l} a_0^{(1)} = 1/4, \quad a_0^{(2)} = 3/4, \quad \alpha^{(1)} = 6\rho^2/25, \quad \alpha^{(2)} = -6\rho^2/25, \\ a_2^{(i)} = 6\alpha^{(i)}, \quad a_1 = -\dfrac{6\rho}{5}, \quad b = -\dfrac{25}{144\rho^2} \end{array}\right\}. \tag{11}$$

Hence, both solutions for $a_1$ and $b$ match, that is, $a_1^{(1)} = a_1^{(2)} = a_1$. Equations (8), (10), and (11) yield the following two solutions to Eq. (6)

$$\psi_1(\xi) = \frac{1}{4}\left(1 + 2\tanh w + \tanh^2 w\right), \tag{12}$$

and

$$\psi_2(\xi) = \frac{1}{4}\left(3 + 2\tanh w - \tanh^2 w\right), \tag{13}$$

where $w = 5\xi/(12\rho)$. These functions were obtained both analytically and numerically [10], and they represent kink solitons. The functions $\psi_1(\xi)$ and $\psi_2(\xi)$ correspond to the positive and negative $\alpha$, respectively. Equation (7) can be written as

$$\alpha = \frac{m\kappa^2}{A}\left(V^2 - c^2\right), \qquad c^2 = l^2\left(k - Kh^2\right)/m, \tag{14}$$

where $V = \omega/\kappa$ and $c$ are the solitonic and linear sound velocities, respectively. As a result, $\psi_1$ describes the supersonic kink, while $\psi_2$ corresponds to the subsonic one.

If we neglect viscosity, that is for $\rho = 0$, the aforementioned parameters can only be expressed through $a_2^{(i)}$ [10]. Hence, $a_2^{(1)}$ and $a_2^{(2)}$ are free parameters. Using Mathematica as above, we obtain [10]

$$\left.\begin{array}{l} a_0^{(1)} = -1/2, \quad a_0^{(2)} = 3/2, \quad a_1 = 0, \\ b^{(1)} = -\dfrac{3}{2a_2^{(1)}}, \quad b^{(2)} = \dfrac{3}{2a_2^{(2)}}, \quad \alpha^{(i)} = a_2^{(i)}/6, \quad i = 1,2 \end{array}\right\}, \tag{15}$$

which yields

$$\psi_{10}(\xi) = \frac{1}{2}\left[-1 + 3\tanh^2\left(\frac{\xi}{2\sqrt{\alpha}}\right)\right], \qquad \alpha > 0, \qquad (16)$$

and

$$\psi_{20}(\xi) = \frac{3}{2}\left[1 - \tanh^2\left(\frac{\xi}{2\sqrt{|\alpha|}}\right)\right], \qquad \alpha < 0. \qquad (17)$$

The solutions $\psi_{10}(\xi)$ and $\psi_{20}(\xi)$ represent the dark and bright solitons, respectively.

### 3. Stability analysis

A purpose of this section is to study stability of the soluitons given by Eqs. (12), (13), (16), and (17). Let us study the solutions $\psi_1(\xi)$ and $\psi_2(\xi)$ first. We introduce the substitution

$$\psi = \psi_i + f, \qquad i = 1, 2 \qquad (18)$$

into Eq. (6). Denoting $f' = F$, we get the following system of the first-order differential equations

$$\left.\begin{array}{l} \dfrac{\mathrm{d}f}{\mathrm{d}\xi} = F \\[2mm] \dfrac{\mathrm{d}F}{\mathrm{d}\xi} = \dfrac{\rho}{\alpha}F - \dfrac{1}{\alpha}f + \dfrac{1}{\alpha}f^2 + \dfrac{2}{\alpha}\psi_i f \end{array}\right\}. \qquad (19)$$

In what follows, we keep only linear terms, as they determine stability. For $\psi_i = \psi_1$, the linear term in $\psi_1 f$ is $\frac{1}{4}f$ and Eq. (19) becomes

$$\left.\begin{array}{l} \dfrac{\mathrm{d}f}{\mathrm{d}\xi} = F \\[2mm] \dfrac{\mathrm{d}F}{\mathrm{d}\xi} = \dfrac{\rho}{\alpha}F - \dfrac{1}{2\alpha}f \end{array}\right\}. \qquad (20)$$

The corresponding eigenvalues can be determined from

$$\begin{vmatrix} -\lambda & 1 \\ -\dfrac{1}{2\alpha} & \dfrac{\rho}{\alpha} - \lambda \end{vmatrix} = 0, \tag{21}$$

and they are

$$\lambda_{1,2} = \frac{\rho \pm \sqrt{\rho^2 - 2\alpha}}{2\alpha}. \tag{22}$$

A solution is stable if $\mathrm{Re}\,(\lambda_{1,2}) < 0$. Keeping in mind that $\rho > 0$ and $\alpha > 0$ for $\psi_0 = \psi_1$, we conclude that $\psi_1(\xi)$ is an unstable solution. The same procedure shows that $\psi_2(\xi)$ is stable. Namely, the eigenvalues are

$$\lambda_{1,2} = \frac{\rho \pm \sqrt{\rho^2 + 2\alpha}}{2\alpha}, \tag{23}$$

and, as $\alpha < 0$, we conclude that $\psi_2(\xi)$ is a stable solution. Therefore, we come to the important conclusion that the subsonic soliton is stable ($\alpha < 0$), while the supersonic one ($\alpha > 0$) is not.

Now, we are going to study the stability of the solutions given by Eqs. (16) and (17), that is, when viscosity is neglected. We put $\rho = 0$ in Eq. (19) and keep only linear terms. This brings about

$$\lambda_{1,2} = \pm i\sqrt{2/|\alpha|}, \tag{24}$$

which means that the linear terms cannot determine whether the solutions are stable or not. In other words, we should use non-linear terms in Eq. (19). Of course, $i$ in Eq. (19) stands for 10 and 20. As

$$\psi_{10}f = -\frac{1}{2}f + \frac{3}{2}f\tanh^2\left(\frac{\xi}{2\sqrt{\alpha}}\right) \tag{25}$$

and

$$\psi_{20}f = \frac{3}{2}f + \frac{3}{2}f\tanh^2\left(\frac{\xi}{2\sqrt{|\alpha|}}\right) \tag{26}$$

we straightforwardly derive

$$\begin{aligned}\frac{dX}{d\xi} &= -\sqrt{\frac{2}{|\alpha|}}\,Y + \frac{1}{\sqrt{2|\alpha|}}\left[Y^2\,\mathrm{sgn}\,\alpha + 3Y\tanh^2\left(\frac{\xi}{2\sqrt{|\alpha|}}\right)\right]\\[2mm]\frac{dY}{d\xi} &= \sqrt{\frac{2}{|\alpha|}}\,X\end{aligned}\Bigg\}. \qquad (27)$$

The functions $X$ and $Y$ were introduced for convenience through

$$f = Y, \qquad F = \sqrt{\frac{2}{|\alpha|}}\,X. \qquad (28)$$

The system (27) cannot be solved analytically. Using the software Mathematica, we obtained $X(\xi)$ and $Y(\xi)$, as shown in Figs. 1 through 7. A positive $\alpha$ refers to $\psi_{10}(\xi)$, while a negative one determines the stability of $\psi_{20}(\xi)$. For Fig. 1, we picked $\alpha = 0.8$, while the initial conditions are $X_0 = Y_0 = 10^{-4}$. We notice a periodic solution. A key point is that a loop in the figure (c) does not circle around the point (0,0), and we conclude that $\psi_{10}$ is not a stable solution. To see this better, we can plot the same figure using $\alpha = 0.8$ and $X_0 = Y_0 = 0.01$. This is shown in Fig. 2.

Let us see what happens for different values of $\alpha$. Figs. 3 and 4 were plotted for $\alpha = 1.5$ and 0.5, respectively. The figures 1, 3, and 4 show that only frequency has been changed, as a higher $\alpha$ corresponds to a lower frequency. Of course, the conclusion regarding stability has been unchanged.

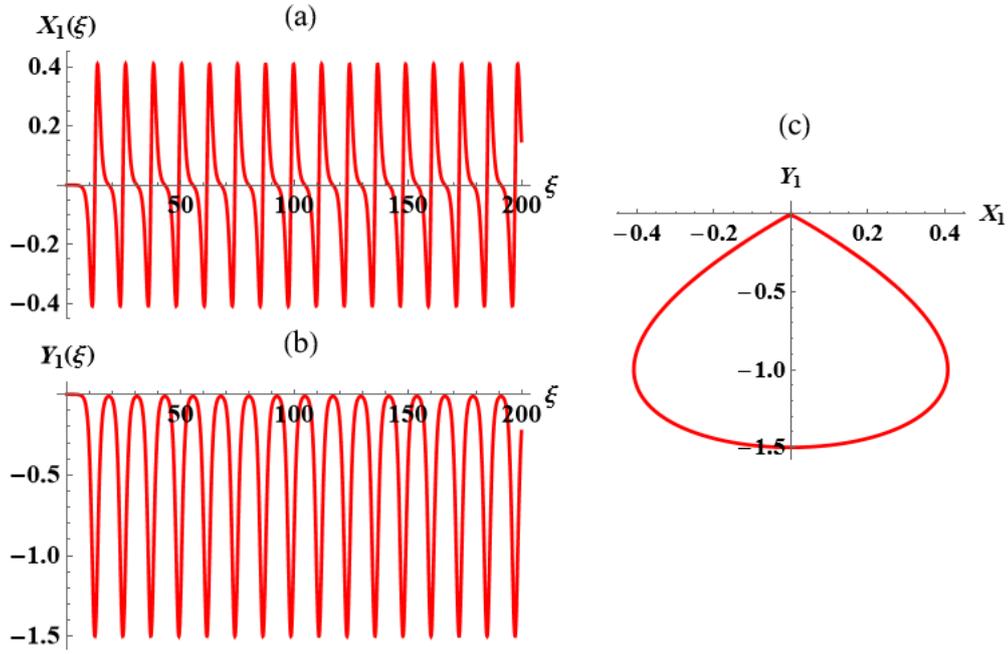

**Fig. 1.** The functions $X_1(\xi)$ (a), $Y_1(\xi)$ (b), and $Y_1(X_1)$ (c), for $\alpha = 0.8$ and $X_0 = Y_0 = 10^{-4}$

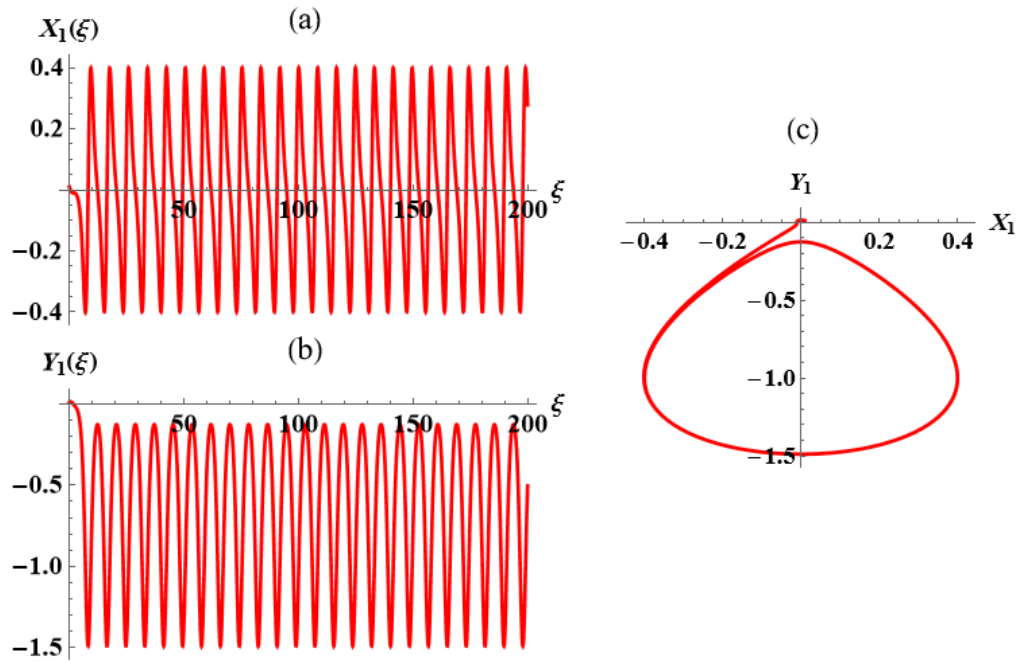

**Fig. 2.** The functions $X_1(\xi)$ (a), $Y_1(\xi)$ (b), and $Y_1(X_1)$ (c), for $\alpha = 0.8$ and $X_0 = Y_0 = 0.01$

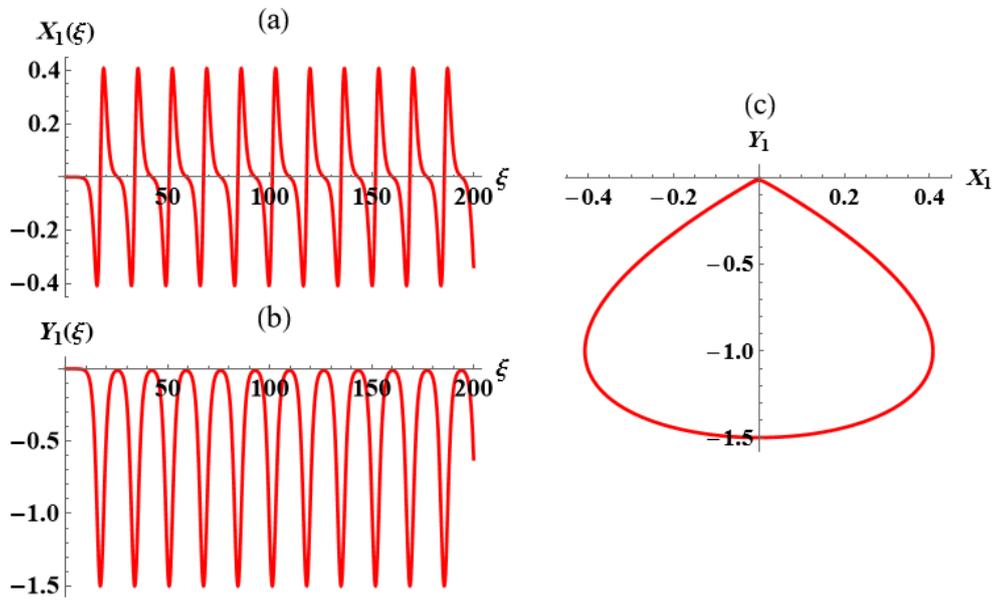

**Fig. 3.** The functions $X_1(\xi)$ (a), $Y_1(\xi)$ (b), and $Y_1(X_1)$ (c), for $\alpha = 1.5$ and $X_0 = Y_0 = 10^{-4}$

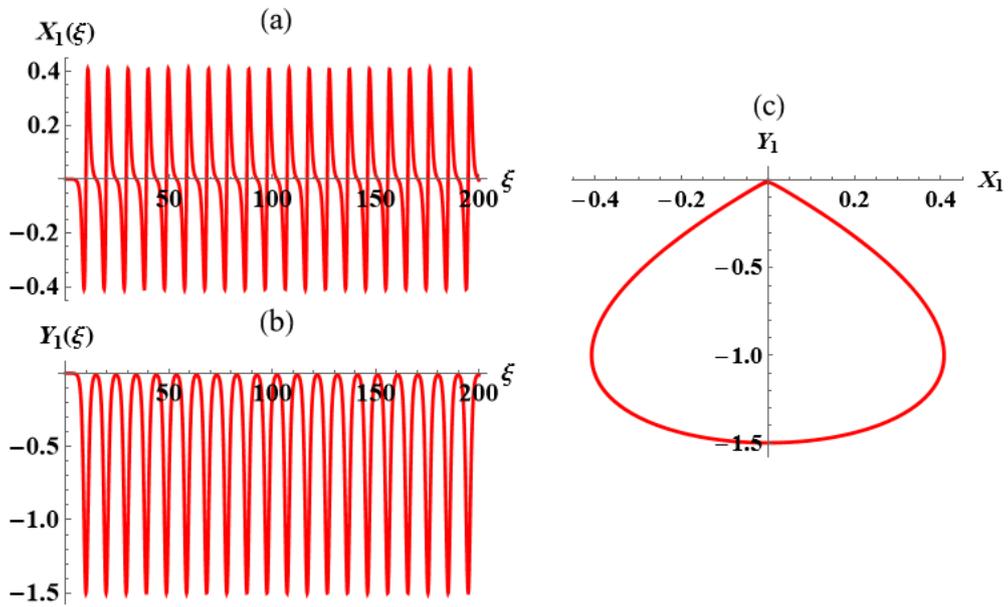

**Fig. 4.** The functions $X_1(\xi)$ (a), $Y_1(\xi)$ (b), and $Y_1(X_1)$ (c), for $\alpha = 0.5$ and $X_0 = Y_0 = 10^{-4}$

It is very interesting to start with negative initial values. For $\alpha = 0.8$ and $X_0 = Y_0 = -0.0001$, we obtain Fig. 5. The solution around zero is not periodic any more. Instead, $X$ and $Y$ reach infinite values for $\xi \approx 12$. Of course, the only conclusion is that this solution is unstable. Notice that a sign of $Y_0$ is only important as the system (27) is symmetrical with respect to x.

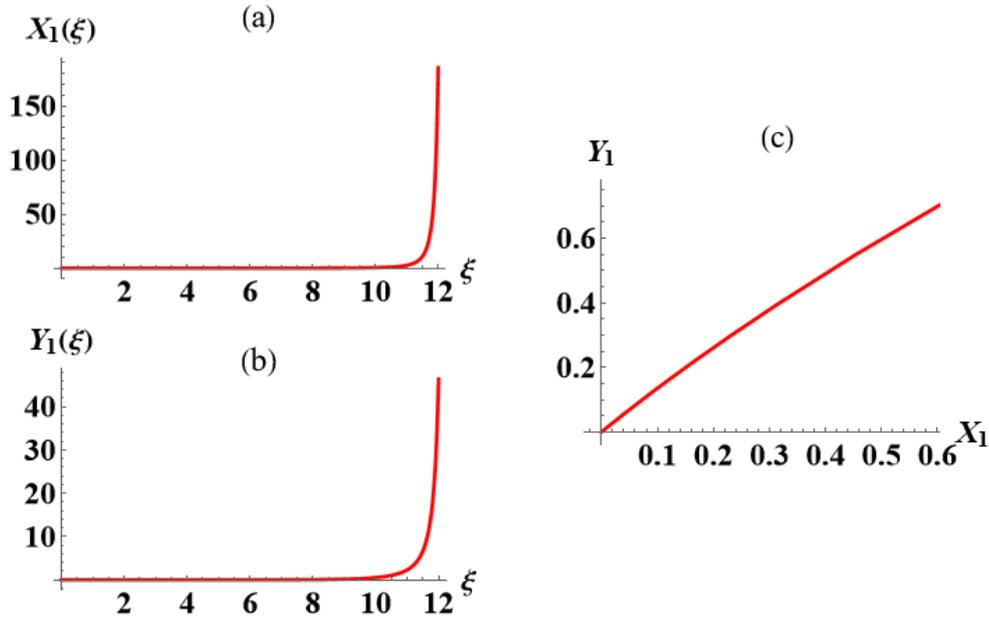

**Fig. 5.** The functions $X_1(\xi)$ (a), $Y_1(\xi)$ (b), and $Y_1(X_1)$ (c), for $\alpha = 0.8$ and $X_0 = Y_0 = -10^{-4}$

Figures 6 and 7 are devoted to the solution $\psi_{20}(\xi)$. Hence, $\alpha$ is negative, and we picked $\alpha = -0.8$. The initial conditions are $X_0 = Y_0 = 10^{-4}$ and $X_0 = Y_0 = -10^{-4}$ for Figs. 6 and 7, respectively. Like above, the only conclusion is that the solution $\psi_{20}(\xi)$ is unstable.

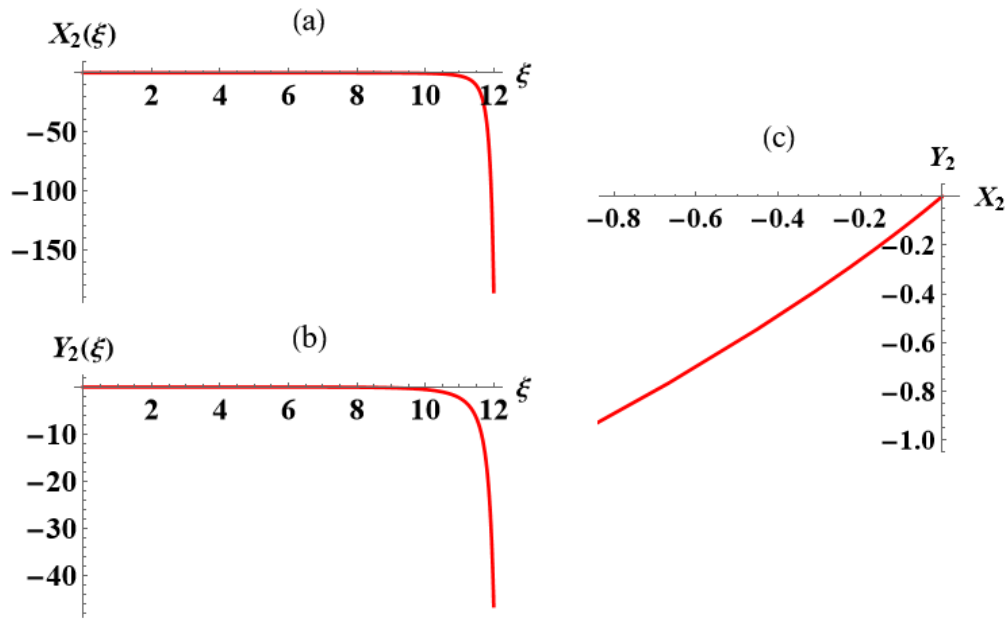

**Fig. 6.** The functions $X_2(\xi)$ (a), $Y_2(\xi)$ (b), and $Y_2(X_2)$ (c), for $\alpha = -0.8$ and $X_0 = Y_0 = 10^{-4}$

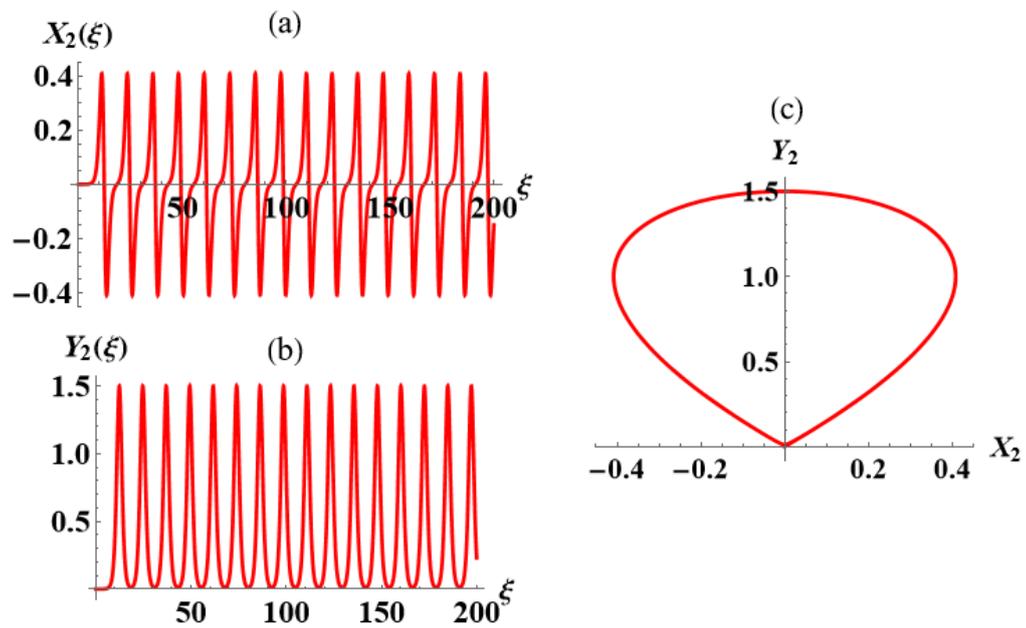

**Fig. 7.** The functions $X_2(\xi)$ (a), $Y_2(\xi)$ (b), and $Y_2(X_2)$ (c), for $\alpha = -0.8$ and $X_0 = Y_0 = -10^{-4}$

## 4. Conclusions

We want to point out on two interesting results. First of all, there are not any stable solutions if viscosity is neglected. From the physics point of view, this probably means that a medium is crucial for the solitary waves to move along MTs. If the frictional forces is taken into consideration, then we obtain, theoretically, both subsonic and supersonic kinks. We showed that only subsonic waves can be stable.